\documentclass[twocolumn,superscriptaddress,longbibliography,
	aps,pra,preprintnumbers]{revtex4-2}

\usepackage[normalem]{ulem}
\usepackage{graphicx}
\usepackage{bm}
\usepackage{color}
\usepackage{epstopdf}
\usepackage{amsmath}
\usepackage{amssymb}
\usepackage{epstopdf}
\usepackage{lipsum}

\usepackage{gensymb}

\usepackage{multirow}

\usepackage{scrextend}

\usepackage[urlcolor=blue,colorlinks=true,citecolor=blue,linkcolor=blue,pdfstartview={FitH},bookmarks=false]{hyperref}

\newcommand{\evec}{\text{e}}

\graphicspath{{fig/}{./fig/}{.}}

\begin{document}

\title{
Chiral phonons in honeycomb sublattice of layered CoSn--like compounds
}

\author{Andrzej~Ptok}
\email[e-mail: ]{aptok@mmj.pl}
\affiliation{\mbox{Institute of Nuclear Physics, Polish Academy of Sciences, W. E. Radzikowskiego 152, PL-31342 Krak\'{o}w, Poland}}

\author{Aksel~Kobia\l{}ka}
%\email[e-mail: ]{akob@kft.umcs.lublin.pl}
\affiliation{\mbox{Institute of Physics, Maria Curie-Sk\l{}odowska University, Plac Marii Sk\l{}odowskiej-Curie 1, PL-20031 Lublin, Poland}}

\author{Ma\l{}gorzata~Sternik}
%\email[e-mail: ]{sternik@wolf.ifj.edu.pl}
\affiliation{\mbox{Institute of Nuclear Physics, Polish Academy of Sciences, W. E. Radzikowskiego 152, PL-31342 Krak\'{o}w, Poland}}

\author{Jan~\L{}a\.{z}ewski}
%\email[e-mail: ]{lazewski@wolf.ifj.edu.pl}
\affiliation{\mbox{Institute of Nuclear Physics, Polish Academy of Sciences, W. E. Radzikowskiego 152, PL-31342 Krak\'{o}w, Poland}}

\author{Pawe\l{}~T.~Jochym}
%\email[e-mail: ]{jochym@wolf.ifj.edu.pl}
\affiliation{\mbox{Institute of Nuclear Physics, Polish Academy of Sciences, W. E. Radzikowskiego 152, PL-31342 Krak\'{o}w, Poland}}

\author{Andrzej~M.~Ole\'{s}}
%\email[e-mail: ]{a.m.oles@fkf.mpg.de}
\affiliation{\mbox{Institute of Theoretical Physics, Jagiellonian University,
Profesora Stanis\l{}awa \L{}ojasiewicza 11, PL-30348 Krak\'{o}w, Poland}}
\affiliation{Max Planck Institute for Solid State Research,
Heisenbergstrasse 1, D-70569 Stuttgart, Germany}

\author{Svetoslav~Stankov}
%\email{svetoslav.stankov@kit.edu}
\affiliation{\mbox{Laboratory for Applications of Synchrotron Radiation, Karlsruhe Institute of Technology, D-76131 Karlsruhe, Germany}}
\affiliation{Institute for Photon Science and Synchrotron Radiation, Karlsruhe Institute of Technology, D-76344 Eggenstein-Leopoldshafen, Germany}

\author{Przemys\l{}aw~Piekarz}
\email[e-mail: ]{piekarz@wolf.ifj.edu.pl}
\affiliation{\mbox{Institute of Nuclear Physics, Polish Academy of Sciences, W. E. Radzikowskiego 152, PL-31342 Krak\'{o}w, Poland}}

\date{\today}

\begin{abstract}
Hexagonal and kagome lattices exhibit extraordinary electronic properties.
It is a natural consequence of additional discrete degree of freedom associated with a valley or the occurence of electronic flat--bands.
Combination of both types of lattices, observed in CoSn-like compounds, leads not only to the topological electronic behavior, but also to the emergence of chiral phonon modes.
Here, we study CoSn-like compounds in the context of realization of chiral phonons.
Previous theoretical studies demonstrated that the chiral phonons can be found in ideal two-dimensional hexagonal or kagome lattices.
However, it turns out that in the case of CoSn-like systems with the $P6/mmm$ symmetry, the kagome lattice formed by {\it d}-block element is decorated by the additional {\it p}-block atom. 
This results in a two dimensional triangular lattice of atoms with non-equal masses and the absence of chiral phonons in the kagome plane.
Contrary to this, the interlayer hexagonal lattice of {\it p}-block atoms is preserved and allows for the realization of chiral phonons.
We discuss properties of these chiral phonons in seven CoSn-like compounds and demonstrate that they do not depend on atomic mass ratio or the presence of intrinsic magnetic order.
The chiral phonons of {\it d}-block atoms can be restored by removing the inversion symmetry.
The latter is possible in the crystal structure of CoGe and RhPb with the reduced symmetry ($P\bar{6}2m$) and distorted-kagome-like lattice.
\end{abstract}

\maketitle

\section{Introduction}
\label{sec.intro}

Graphene -- an exact, two-dimensional (2D) honeycomb lattice, exhibits a range of extraordinary electronic properties~\cite{novoselov.geim.05,castroneto.guinea.09,peres.10,dassarma.adam.11,goerbig.11,koto.uchoa.12}.
The existence of two degenerate and inequivalent valleys at the corners of the Brillouin zone (BZ) constitutes an additional discrete degree of freedom.
This leads to a proposal of the \textit{valleytronics}~\cite{bussolotti.kawai.18}, which contrary to spintronics concept~\cite{avsar.ochoa.20}, manipulates a valley index instead of carrier spin.
Recently, the potential application in valleytronics for several three-dimensional (3D) systems (e.g.\ transition metal dichalcogenides) has been suggested~\cite{schaibley.yu.16,wang.chernikov.18,liu.gao.19}.

Due to the broken inversion symmetry of crystal structure, electrons in both valleys 
experience an effective magnetic field with equal magnitudes but opposite signs~\cite{xiao.yao.07}.
Such behavior, associated with effective angular momenta at the point with opposite valley index, opens a way for optical pumping of valley-polarized carriers by circularly polarized light~\cite{cao.wang.12,mak.he.12,sallen.bouet.12,zeng.dai.12,wu.ross.13}. 
Additionally, new phenomena like valley Hall effect can be realized~\cite{xiao.liu.12,mak.mcgill.14,tong.gong.16,jin.kim.18}, which is analogue to the spin Hall effect~\cite{sinova.valenzuela.15}.
The mentioned transport properties exhibit a topological nature, which is evidenced by the presence of the finite Berry curvature in the system~\cite{xiao.chang.10,hasan.kane.10,bansil.lin.16}.

Interestingly, electronic systems are not unique in presenting such topological properties, as this is possible also for various types of bosonic systems~\cite{liu.chen.20,mcclarty.21}.
The topological bosons are not an ordinary extension of the topological fermions due to significant difference between both systems (e.g.\ different types of statistics and interactions).
However, from theoretical point of view, both of them can be described by the Bloch wave function in periodic lattices~\cite{bloch.29}.
The topological concepts widely used in electronic systems, such as Berry phase, Berry curvature, or Chern number, are applicable to bosonic systems as well.

As an example, a topologically nontrivial phase was observed in photonic
systems~\cite{lu.joannopoulos.14,noh.huang.17,ozawa.price.19}.
Recently, many other topological features were reported in these 
types of systems, e.g.\ Weyl points~\cite{lu.fu.13,lin.xiao.16,hou.chen.16,he.qiu.18,chen.xiao.16}, topological edge modes~\cite{chen.xiao.16,noh.huang.17,chen.zhao.17,phan.liu.21} or a topological acoustic wave in metamaterial systems~\cite{yang.gao.15,xiao.chen.15,lu.qiu.17,li.qiu.17,he.qiu.18,zhang.cheng.18}.

Stemming from that, topological properties can be also expected in the case of phononic systems~\cite{li.ren.12,li.liu.21}.
The topological properties of phonons can be a consequence of acoustic-- and optical--mode inversion in phonon dispersion relations~\cite{singh.wu.18}, that is an exact analogue of the band inversion in the electronic structure.
Consequently, also Dirac and/or Weyl high-degenerated points~\cite{jin.wang.18,zhang.song.18,miao.zhang.18,lie.xie.18,xie.li.19,liu.hou.19,zhong.liu.21,liu.qian.20,peng.hu.20,chen.wang.21,liu.li.21,liu.wang.21,li.xie.20,he.rivera.20} as well as nodal lines~\cite{he.rivera.20,li.xie.20,stenull.kane.16} can occur.

% exactly circular

One of the possible examples of topological phenomena concerning phonons is the ability to distinguish them via the chiral modes~\cite{chen.zhang.18}. 
These types of modes are characterized by the opposite chirality and can emerge in the system due to the quantized Berry phase and pseudoangular momenta~\cite{zhang.niu.14,liu.lian.17,zhang.niu.15}.
In analogy to the electronic structure of graphene, in the phonon spectrum of the hexagonal lattice the opposite Berry phase in the K and K' points are observed~\cite{liu.xu.17,zhang.niu.15,jin.wang.18,li.wang.20}.
The existence of chiral phonons allows for the realization of the elastic analogue of the valley Hall effect~\cite{pal.ruzzene.17}.
Chiral phonons have been reported also in the kagome lattice~\cite{chen.wu.19}, in the bilayer triangular~\cite{xu.wei.18} or bilayer hexagonal~\cite{gao.zhang.18} lattices.
Such type of phonons was also recently observed in the transition metal dichalcogenide WSe$_{2}$~\cite{zhu.yi.18} and predicted to exist in a MoS$_{2}$/WS$_{2}$ heterostructure~\cite{zhang.srivastava.20}.
Additionally, recent studies show the possibility of experimental realization of the chiral phonons in more complex systems like BiB$_{3}$O$_{6}$~\cite{romao.19}.

%%%%%%%%%%%%%%%%%%%%
%%%%%%%%%%%%%%%%%%%%
%%%%%%%%%%%%%%%%%%%%

\begin{figure}[!t]
\centering
\includegraphics[width=\linewidth]{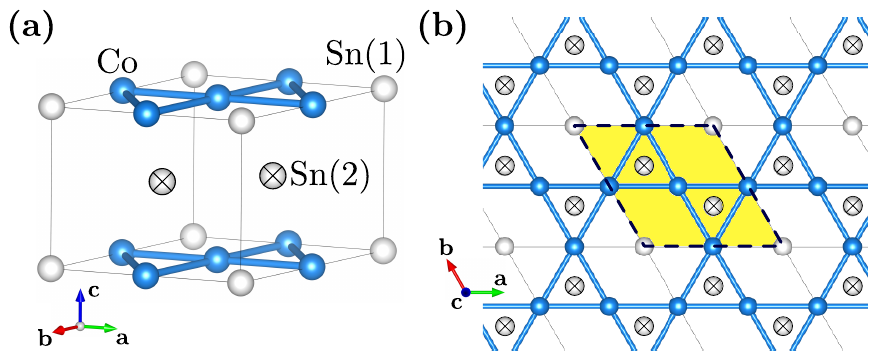}
\caption{
(a) Side- and (b) top view of the CoSn-like crystal structure with {\it P6/mmm} symmetry.
The Co atoms (blue balls) form the kagome lattice, while the Sn atoms occupy two non--equivalent positions: Sn(1) located in the plane of the kagome lattice (grey balls), and Sn(2) located between two kagome-lattice planes and forming a honeycomb sublattice (crossed balls).
\label{fig.crystal}
}
\end{figure}

{\it Motivation.}
Several groups of 3D crystals can combine into two different types of 2D topological lattices.
One such example is the structure of CoSn-like compounds [cf.\ Fig.~\ref{fig.crystal}(a)], formed by $d$-block element (like Fe, Co, Ni, Rh, or Pt) and $p$-block element (like Ge, In, Sn, Tl, or Pb).
These compounds crystallize within the $P6/mmm$ symmetry --- the $d$-block elements form a kagome sublattice, while the $p$-block elements have two nonequivalent positions: one position in the plane of the kagome lattice, and second intercollated between two kagome-lattice planes [marked in Fig.~\ref{fig.crystal} as Sn(1) and Sn(2), respectively].
Atoms in Sn(2)--type position form the honeycomb sublattice [atoms marked by cross in Fig.~\ref{fig.crystal}(b)].
This complex structure arises due to existence of the kagome lattice and allows for a realization of topological nearly-flat band in electronic band structure~\cite{meier.du.20,liu.li.20}.

In our study we investigate the following compounds: CoSn, CoGe, FeSn, FeGe, NiIn, RhPb, and PtTl.
These materials permit a systematic study of the impact of various parameters e.g., atomic mass ratio or magnetic order on the chiral phonons in real 3D systems. 
Additionally, the application of an external uniaxial pressure on the system can mimick the effective magnetic field or the spin--orbit coupling-like effects acting on phonons.
Furthermore, we found that two of the investigated compounds (i.e. CoGe and RhPb) cannot exist in the $P6/mmm$ structure. 
Instead they exhibit $P\bar{6}2m$ symmetry, what was not reported previously in literature.
Our results suggests a possible emergence of chiral phonon also in these structures.
Summarizing, we systematically study and discuss the origin of the chiral phonons in a large class of real 3D systems.

The paper is organized as follows.
Details of the {\it ab initio} calculations are present in Sec~\ref{sec.theo}.
Next, in Sec.~\ref{sec.res}, we discuss the numerical results.
Possible experimental consequences of realization the chiral phonons are discussed in Sec.~\ref{sec.exp}.
Finally, a summary is included in Sec.~\ref{sec.sum}.

%%%%%%%%%%%%%%%%%%%%%%%%%%%%%%%%%%%%%%%%%%
%%%%%%%%%%%%%%%%%%%%%%%%%%%%%%%%%%%%%%%%%%
%%%%%%%%%%%%%%%%%%%%%%%%%%%%%%%%%%%%%%%%%%

\section{Theoretical background}
\label{sec.theo}

\subsection{Numerical calculation details}
\label{sec.num_det}

The first-principles (DFT) calculations are preformed using the projector augmented-wave (PAW) potentials~\cite{blochl.94} implemented in the Vienna Ab initio Simulation Package ({\sc Vasp}) code~\cite{kresse.hafner.94,kresse.furthmuller.96,kresse.joubert.99}.
The calculations are made within generalized gradient approximation (GGA) in the Perdew, Burke, and Ernzerhof (PBE) parametrization~\cite{pardew.burke.96}.
The energy cutoff for the plane-wave expansion was set to $350$~eV.
The calculations carried out without and with spin polarization in a system with different spin configurations allow to determine the magnetic ground state.
Optimizations of the structural parameters (lattice constants and atomic positions) for nonmagnetic and ferromagnetic order are performed in the primitive unit cell using $10 \times 10 \times 6$ {\bf k}--point grid in the Monkhorst--Pack scheme~\cite{monkhorst.pack.76}. For the antiferromagnetic structure the doubled unit cell and the reduced {\bf k}--point grid ($10 \times 10 \times 3$) was used.
As a break of the optimization loop, we take the condition with an energy difference of $10^{-5}$~eV and $10^{-7}$~eV for ionic and electronic degrees of freedom.

The interatomic force constants (IFC) are calculated by {\sc Alamode} software~\cite{tadano.gohda.14}, using the $2 \times 2 \times 2$ supercell with $48$ atoms.
Calculations are performed for the thermal distribution of multi-displacement of atoms at $T = 50$~K, generated within {\sc hecss} procedure ~\cite{jochym.lazewski.21}.
The energy and the Hellmann-Feynman forces acting on all atoms are calculated with {\sc Vasp} for one hundred different configurations of atomic displacements in the supercell.
In dynamical properties calculations, we include first- and second-orders phonon contributions, which correspond to harmonic and cubic IFC, respectively.

\subsection{Dynamical matrix and polarization vector}
\label{sec.dym_mat}

The lattice dynamics of the system can be studied by the dynamical matrix:
\begin{eqnarray}
\nonumber D_{\alpha\beta}^{jj'} \left( {\bm q} \right) \equiv \frac{1}{\sqrt{m_{j}m_{j'}}} \sum_{n} \Phi_{\alpha\beta} \left( j0, j'n \right) \exp \left( i {\bm q} \cdot {\bm R}_{j'n} \right) , \\
\label{eq.dyn_mat}
\end{eqnarray}
where ${\bm q}$ is the phonon wave vector and $m_{j}$ denotes the mass of $j$th atom.
Here, $\Phi_{\alpha\beta} \left( j0, j'n \right)$ is the IFC tensor ($\alpha$ and $\beta$ denotes the direction index, i.e.\ $x$, $y$, and $z$) between $j$th and $j'$th atoms located in the initial (0) and $n$th primitive unit cell. 
The phonon spectrum for a given wave vector ${\bm q}$ is specified by the eigenproblem of the dynamical matrix~(\ref{eq.dyn_mat}), i.e.:
\begin{eqnarray}
\omega^{2}_{\varepsilon{\bm q}} \evec_{\varepsilon{\bm q}\alpha j} = \sum_{j'\beta} D_{\alpha\beta}^{jj'} \left( {\bm q} \right) \evec_{\varepsilon{\bm q}\beta j'} .
\end{eqnarray}
Here, the $\varepsilon$ branch describes the phonon with $\omega_{\varepsilon{\bm q}}$ frequency and polarization vector $\evec_{\varepsilon{\bm q}\alpha j}$.
Each $\alpha j$ component of the polarization vector denotes displacement of $j$th atom in $\alpha$th direction.

\subsection{Circular phonon polarization}
\label{sec.circ_pol}

In practice, the polarization vector $\evec_{\varepsilon{\bm q}\alpha j}$ is related to the oscillation of each atom caused by the propagation of a $\varepsilon$ phonon with wave vector ${\bm q}$. 
Each $j$th atom is described by three components of $\evec_{\varepsilon{\bm q}\alpha j}$, which refer to independent harmonic oscillators in each of $\alpha$ directions.
The circular polarization occurs when the two orthogonal oscillators are of equal magnitude and are out of phase by exactly $\pm \pi/2$.
To study this behavior, we employ Jones vectors $\frac{1}{\sqrt{2}} ( 1 \; \pm i )^\text{T}$ (denoting circular polarization in $xy$ plane), where upper and lower sign correspond to left-handed polarization (LHP) and right-handed polarization (RHP), respectively.
From the theoretical point of view, Jones vectors act on every $j$th atom, and can be used to introduce a new basis defined as:
$|R_{1}\rangle \equiv \frac{1}{\sqrt{2}} \left( 1 \; i \; 0 \cdots 0 \right)^\text{T}$, $|L_{1}\rangle \equiv \frac{1}{\sqrt{2}} \left( 1 \; -i \; 0 \cdots 0 \right)^\text{T}$, $|Z_{1}\rangle \equiv \left( 0 \; 0 \; 1 \cdots 0 \right)^\text{T}$; $\ldots$; $|R_{j}\rangle \equiv \frac{1}{\sqrt{2}} \left( 0 \cdots 1 \; i \; 0 \cdots 0 \right)^\text{T}$, $|L_{j}\rangle \equiv \frac{1}{\sqrt{2}} \left( 0 \cdots 1 \; -i \; 0 \cdots 0 \right)^\text{T}$, $|Z_{j}\rangle \equiv \left( 0 \cdots 0 \; 0 \; 1 \cdots 0 \right)^\text{T}$; $\ldots$; i.e., two in-plane components of circular oscillation are replaced by the Jones vectors coefficients, while third component is unchanged.
In this basis, each polarization vector $\evec \equiv \evec_{\varepsilon{\bm q}\alpha j}$, can be represented as:
\begin{eqnarray}
\evec = \sum_{j} \left( \alpha^{R}_{j} | R_{j} \rangle + \alpha^{L}_{j} | L_{j} \rangle + \alpha^{Z}_{j} | Z_{j} \rangle \right) ,
\end{eqnarray}
where $\alpha^{V}_{j} = \langle V_{j} | \evec \rangle$, for $V \in \{ R, L, Z \}$ and $j \in \{ 1 , 2, \cdots, N \}$ (N is a total number of atoms in a primitive unit cell).
Additionally, we can define the phonon circular polarization operator $\hat{S}^{z}_{ph}$ along the $z$ direction as:
\begin{eqnarray}
\hat{S}^{z}_{ph} \equiv \sum_{j} \hat{S}_{j}^{z} = \sum_{j} \left( | {R}_{j} \rangle \langle R_{j} | - | L_{j} \rangle \langle L_{j} | \right) ,
\end{eqnarray}
where $\hat{S}_{i}^{z}$ is the phonon circular polarization operator at site $i$.
From this, the phonon circular polarization $s_{j}^{z}$ of $j$th atom can be expressed as
\begin{eqnarray}
s_{j}^{z} = \hslash \evec^{\dagger} \hat{S}_{j}^{z} \evec = \hslash \left( | \alpha^{R}_{j} |^{2} - | \alpha^{L}_{j} |^{2} \right) .
\end{eqnarray}
It corresponds to the phonon angular momentum along the $z$ direction~\cite{mclellan.88,zhang.niu.14}. 
However, in the general case, we can discuss the angular momenta along some arbitrary direction.
For $s_{j}^{z} > 0$ ($s_{j}^{z} < 0$) the phonon mode has RHP (LHP), while for $s_{j}^{z} = 0$ the phonon mode is linearly vibrating.
Finally, the total phonon circular polarization $s_{ph}^{z} = \sum_{j} s_{j}^{z}$ denotes polarization of a whole system.

\section{Results and discussion}
\label{sec.res}

\begin{table}[!b]
\caption{
Comparison of the experimental and theoretical lattice constants for crystals in $P6/mmm$ symmetry.
The results are obtained in the presence of spin-orbit coupling.
$\Theta$~denotes the mass ratio of components, i.e.\ relation of the mass of $d$-block element with respect to the mass of $p$-block element.
}
\begin{ruledtabular}
\begin{tabular}{lccccc}
 & & \multicolumn{2}{c}{Theory} & \multicolumn{2}{c}{Exp.~(Ref.~\cite{meier.du.20})} \\
 & $\Theta$ (--) & a (\AA) & c (\AA) & a (\AA) & c (\AA) \\
 \hline
\multicolumn{6}{c}{ $P6/mmm$ (SG: 191) } \\
 \hline
CoSn & 0.50 & 5.289 & 4.224 & 5.2693 & 4.2431 \\
CoGe~\footnote{\label{tab.1sttablefoot}unstable in $P6/mmm$ structure, cf.\ discussion in Sec.~\ref{sec.res}} & 0.81 & 4.987 & 3.876 & --- & --- \\
FeSn~\footnote{AFM order} & 0.47 & 5.285 & 4.456 & 5.2765 & 4.4443 \\
FeGe~\footnote{FM order} & 0.77 & 4.963 & 4.063 & 4.9852 & 4.0482 \\
NiIn & 0.51 & 5.280 & 4.377 & 5.2296 & 4.3390 \\
RhPb~\footref{tab.1sttablefoot} & 0.49 & 5.740 & 4.487 & 5.6660 & 4.4127 \\
PtTl & 0.96 & 5.702 & 4.797 & 5.6017 & 4.6276 \\
 \hline
\multicolumn{6}{c}{ $P\bar{6}2m$ (SG: 189) } \\
 \hline
CoGe & 0.81 & 5.009 & 3.857 & --- & --- \\
RhPb & 0.49 & 5.770 & 4.464 & --- & ---
\end{tabular}
\end{ruledtabular}
\label{tab.latt}
\end{table}

\subsection{Crystal structure}
\label{sec.crys_struc}

\begin{figure*}[!tp]
\centering
\includegraphics[width=\linewidth]{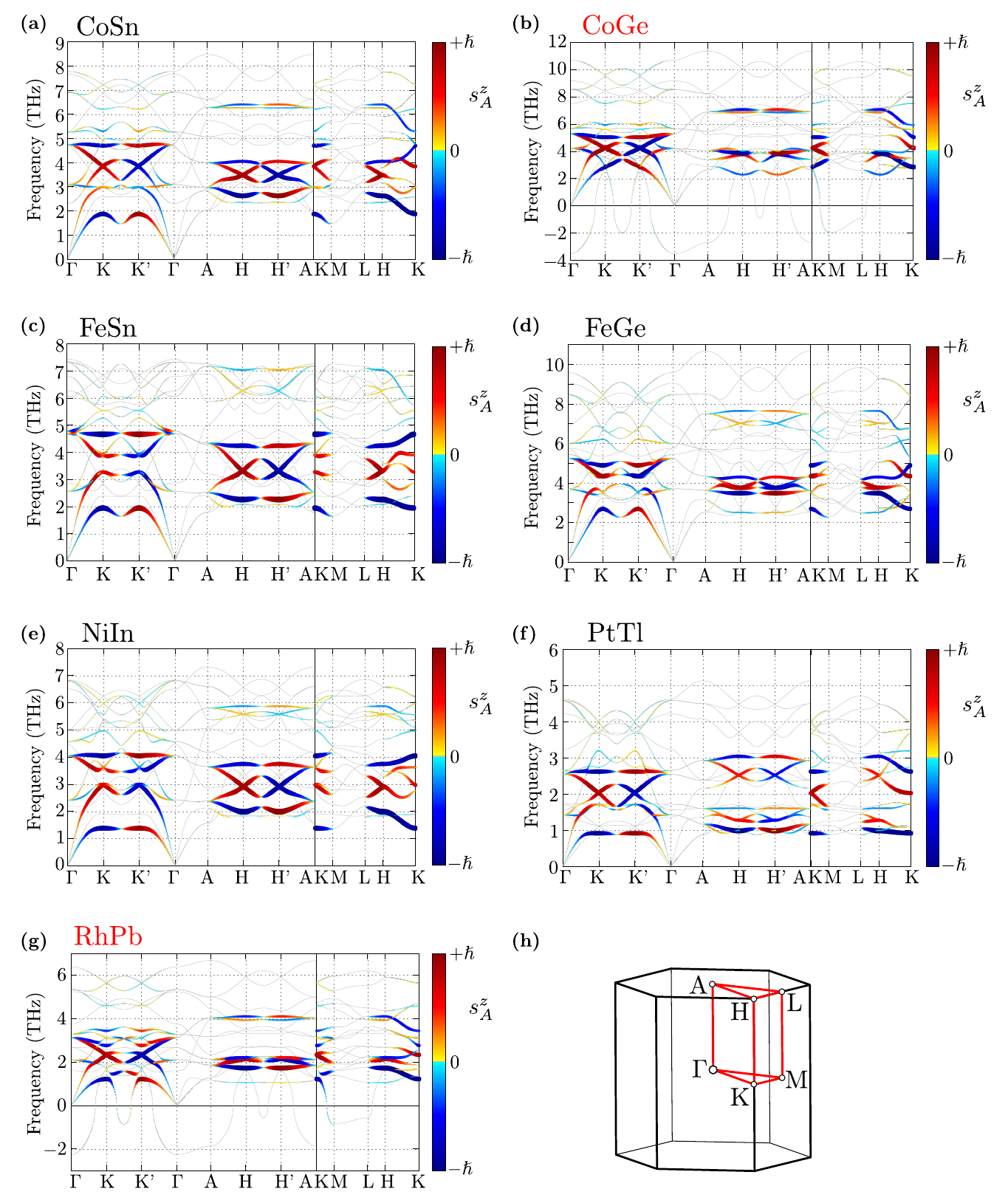}
\caption{
(a)--(g) The phonon dispersions for the studied compounds (as labeled), along the high symmetry points of the first Brillouin zone, presented in (h), of the $P6/mmm$ structure.
The color and width of line correspond to the 
phonon angular momentum of the {\it p}-block element in sublattice A of the honeycomb net (cf.\ Fig.~\ref{fig.chiral_modes}).
Atoms in sublattice B are described by the phonon angular momentum with the opposite value.
Red labels denotes the compounds unstable in the $P6/mmm$ structure.
\label{fig.ban191}
}
\end{figure*}

The CoSn--like compounds typically crystallize in the hexagonal structure with $P6/mmm$ symmetry (space group: 191).
The {\it d}-block element is placed at $3f$ $(1/2, 0, 0)$ Wyckoff position, while the {\it p}-block atoms occupy two non-equivalent $1a$ $(0, 0, 0)$ and $2d$ $(1/3, 2/3, 1/2)$ Wyckoff positions.
Lattice parameters of the optimized ground-state structures are summarized in Table~\ref{tab.latt}. As shown, they remain in close agreement with the available experimental data. 
Two of the seven studied compounds have been previously reported to be magnetically ordered. 
The magnetic moments on Fe atoms are aligned ferromagnetically (FM) in FeGe~\cite{sales.yan.19,sales.meier.21}, while in FeSn the antiferromagnetic (AFM) order is observed~\cite{zeng.kent.06}. 
In AFM structure the spins are arranged ferromagnetically in each FeSn layer and antiferromagnetically ordered in neighboring layers.

For all compounds with the $P6/mmm$ symmetry, the phonon dispersion relations are calculated and presented in Fig.~\ref{fig.ban191}.
The irreducible representations at the $\Gamma$ point are: $A_\text{2u} + E_\text{1u}$ for acoustic modes, and $2 A_\text{2u} + B_\text{1g} + B_\text{1u}+ B_\text{2u} + E_\text{2u} + E_\text{2g} + 3 E_\text{1u}$ for optic modes.
The phonon dispersions are discussed in details in Sec.~\ref{sec.topo_ph}, but here we would like to point out that two crystals, CoGe and RhPb, are not stable in the $P6/mmm$ structure because their phonon spectra exhibit imaginary frequencies.
From analysis of the soft mode with the lowest frequency (at the $\Gamma$ point), we found a stable structure for these two compounds, i.e.\ the $P\bar{6}2m$ hexagonal symmetry (space group 189), for details see Sec.~\ref{sec.sg189}.
For both materials, the lattice constants of a new structure remain almost unchanged (Table~\ref{tab.latt}), however, the {\it d}-block element is slightly shifted from the original high-symmetry site $(0.5, 0, 0)$ in the $P6/mmm$ structure to new $3f$ positions of a $P\bar{6}2m$ space group, $(0.4665, 0, 0)$ or $(0.4669, 0, 0)$ for Co or Rh, respectively.

\begin{figure}[!b]
\centering
\includegraphics[width=\linewidth]{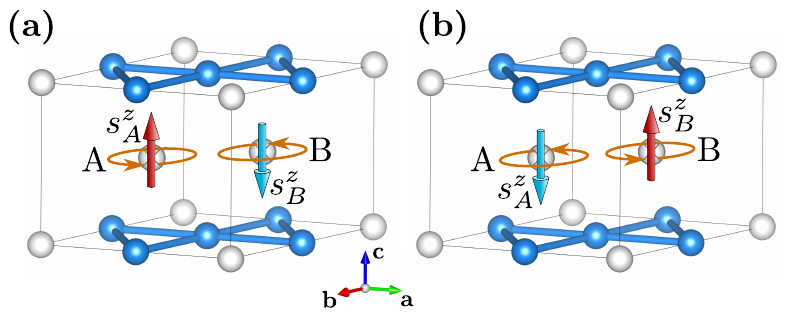}
\caption{
The phonons angular momentum in two non-equivalent sites
of the honeycomb net for the $P6/mmm$ structure. 
The {\it p}-block elements located on sites A and B of honeycomb net have the opposite phonon angular momenta.
\label{fig.chiral_modes}
}
\end{figure}

%%%%%%%%%%%%%%%%%%%%%%%%%%%%%%%%%%%%%%%%%%
%%%%%%%%%%%%%%%%%%%%%%%%%%%%%%%%%%%%%%%%%%

\subsection{Chiral phonons in $\bm{P6/mmm}$ symmetry}
\label{sec.topo_ph}

The phonon dispersion curves of the studied compounds are presented in Fig.~\ref{fig.ban191}.
Each compound realizes phonon flat-bands, which can be the source of 
the collective atomic vibrations in kagome-sublattice plane~\cite{yin.shumiya.20}.
However, analysis of the phonon angular momentum (phonon dispersion curves distinguished by colors and width of lines in Fig.~\ref{fig.ban191}), shows that these bands are mostly associated with chiral phonons realized by the {\it p}-block elements within the honeycomb sublattice (see movies \texttt{k\_mode$\ast$} in the Supplemental Material (SM)~\footnote{
See the Supplemental Material at [URL will be inserted by publisher] for visualizations of the chosen phonon modes.
There, yellow (green) orbs correspond to the {\it p}-block ({\it d}-block) atoms.
Files named \texttt{k\_mode$\ast$} present modes from K point, while \texttt{gamma\_mode4} correspond to the soft mode at $\Gamma$ point in CoGe and RhPb.
}).
Two atoms composing this sublattice, exhibit an opposite phonon angular momentum, what is schematically presented in Fig.~\ref{fig.chiral_modes}.
For example, when atom located at the A site has $s_{A}^{z} = \pm\hslash$, then atom at the B site has $s_{B}^{z} = \mp\hslash$.

Interestingly, chiral modes preserve the time reversal symmetry $\left( s_{j}^{z} , {\bm k} \right) \rightarrow \left( -s_{j}^{z} , -{\bm k} \right)$, i.e.,\ for K and K' points of the reciprocal space, 
since components of polarization vectors for each atom correspond to opposite phonon angular momenta as it is depicted on the left panel for K and right panel for K' of Fig.~\ref{fig.chiral_modes}).
Along the path from K to K' point, the phonon angular momentum changes its value continuously and equals to zero at point M (i.e., precisely half way between K and K').
Similarly, the phonon angular momentum also vanishes at the $\Gamma$ point.

\begin{figure}[!t]
\centering
\includegraphics[width=\linewidth]{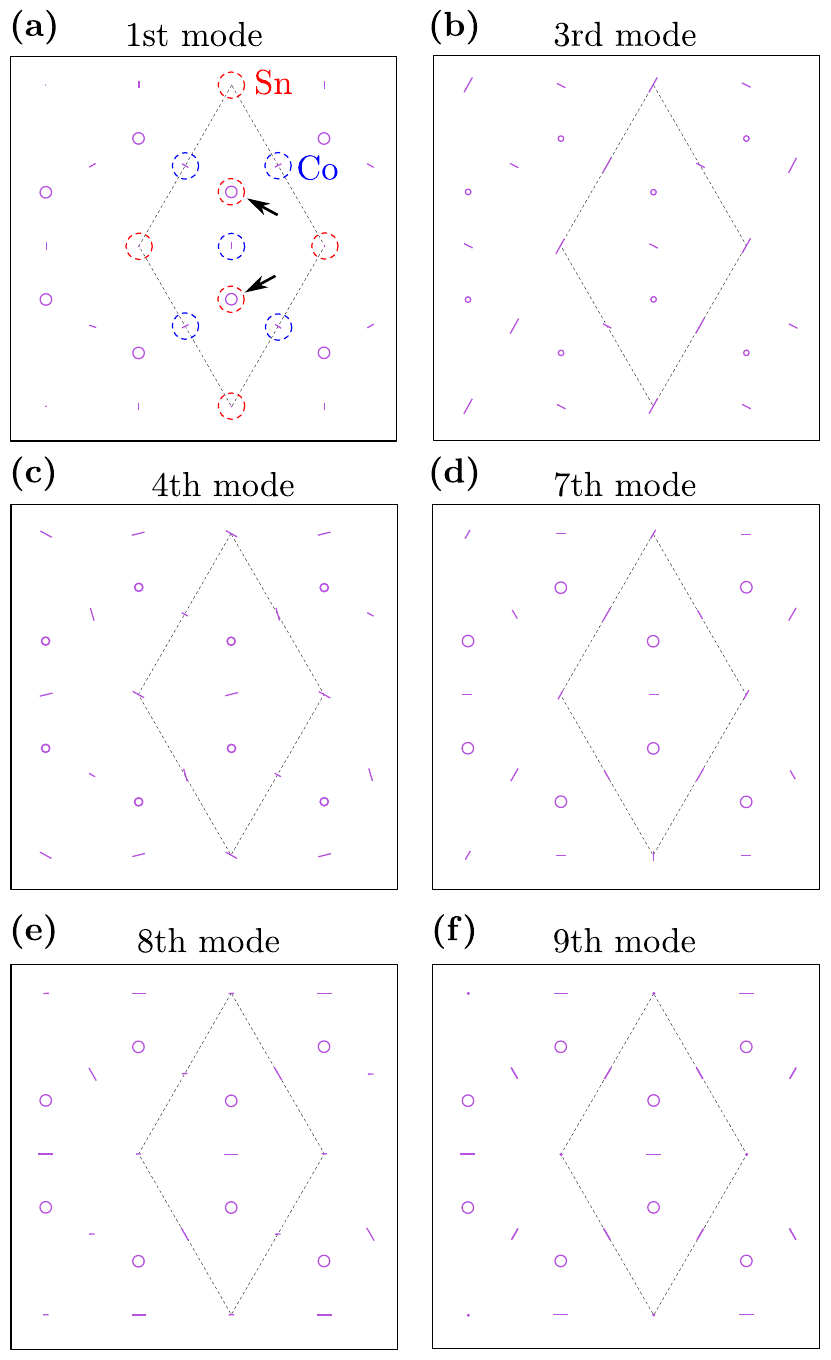}
\caption{
Examples of trajectories (violet lines) of the CoSn atoms (for $P6/mmm$ symmetry), generated by chosen modes at the K point (as labeled).
The dashed rhombus represents the unit cell, while equilibrium position of atoms are marked by the blue (Co atoms) and red (Sn atoms) circles on panel (a).
The black arrows indicate the Sn atoms belonging to the honeycomb lattice and showing the chiral vibrations.
\label{fig.traj}
}
\end{figure}

As we can see in Fig.~\ref{fig.ban191}, the phonon branches with strong phonon angular momentum (marked by thick lines), are located mostly in the mid-frequency range and emerge from the double-degenerated $E_\text{1u}$ modes at the $\Gamma$ point. 
Moving away from the $\Gamma$ point these modes become non-degenerate and lead to the creation of circular-like rotation of {\it p}-block elements in the honeycomb lattice (this also happens at the $\Gamma$ point, however, in this 
case circular polarization results from the composition of two double-degenerated $E_\text{1u}$ or $E_\text{2g}$ modes.
In the high-frequency range of the spectrum, a small but nonzero phonon angular momentum is also observed.
These modes correspond to oscillations in the kagome layer, which also indirectly affect the honeycomb sublattice.
Some examples of atom trajectories generated by the chosen modes are shown in Fig.~\ref{fig.traj}.
When the phonon angular momentum is relatively large (approximately equals to a nominal value $\pm\hslash$), the {\it p}-block atoms realize a full circular movement with a small contribution from atoms in the kagome layer [e.g.~Fig.~\ref{fig.traj}(a)].
On the other hand, a small value of the angular momentum (much smaller than the nominal value), corresponds to a small circular [e.g.~Fig.~\ref{fig.traj}(b)] or elliptic-like oscillations, and also affects {\it d}-block atoms (realizing linear oscillation).
Movies visualizing the presented trajectories are included in the SM~\cite{Note1}.

\begin{figure}[t!]
\centering
\includegraphics[width=0.972\linewidth]{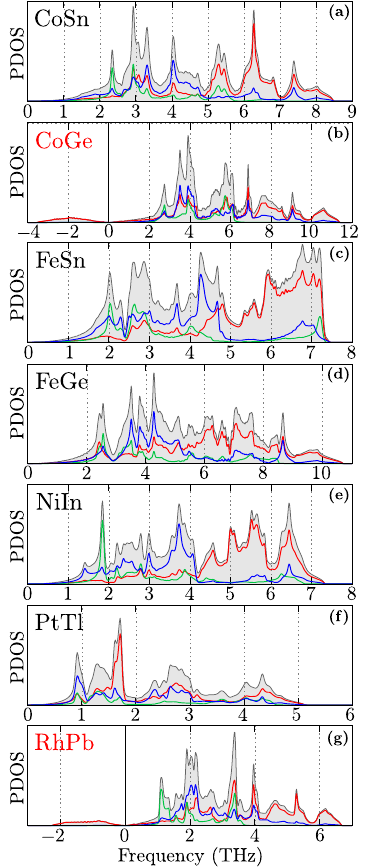}
\caption{
The total and partial phonon density of states (DOS) for the studied compounds (as labeled) of the $P6/mmm$ structure.
The black line (gray area) corresponds to the total DOS, while red, green, and blue lines correspond to the partial DOS for {\it d}-block element (kagome sublattice), {\it p}-block element in kagome plane [position Sn(1) in Fig.~\ref{fig.crystal}], and {\it p}-block elements in the honeycomb sublattice [position Sn(2) in Fig.~\ref{fig.crystal}].
\label{fig.dos191}
}
\end{figure}

The chiral phonons are not observed in the kagome layer of the {\it d}-block atoms.
In fact, this layer is composed of the kagome sublattice and one {\it p}-block atom forming together a two dimensional triangular lattice of atoms with non-equal masses.
Nevertheless, the {\it p}-block atoms within the honeycomb net can form a circular movement with relatively ``small'' nonzero phonon angular momentum, even if the modes propagate within the kagome layer. 
These features are clearly visible in the case of high-frequency modes propagating with wavevectors H or H' (see Fig.~\ref{fig.ban191}).
In such a case, the observed modes realized mostly vibration of {\it d}-block atoms with a small contribution of {\it p}-block atoms vibration.

Phonon dispersion relations also exhibit nodal lines (along K--H or K'--H' paths), which are allowed by the discussed space group and their \textit{build--in} $C_{3v}$ symmetries~\cite{liu.jin.21}.
Some of these nodal lines are composed of fourfold degenerate branches with strong phonon angular momentum (characteristic blue or red $\mathsf{X}$-cut of two branches at K or K' point). This suggests the possibility for the realization of a phonon ballistic transport in some situations.

\begin{figure*}[!t]
\centering
\includegraphics[width=\linewidth]{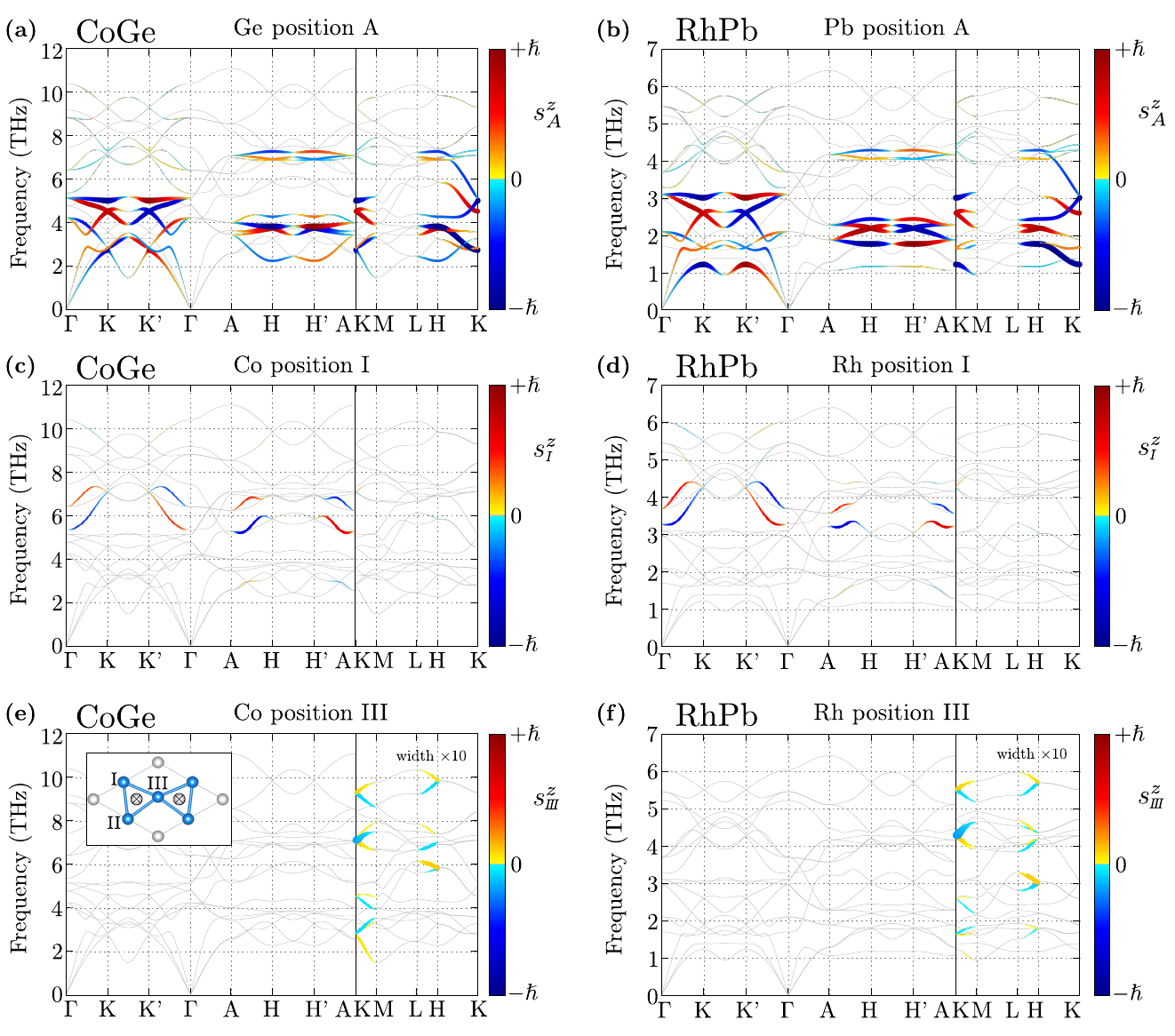}
\caption{
Phonon dispersions of CoGe and RhPb in the stable $P\bar{6}2m$ structure.
Panels from top to bottom represent the phonon angular momentum of the {\it p}-block element in the A sublattice  of the honeycomb net (cf.~Fig.~\ref{fig.chiral_modes}), and the {\it d}-block atoms in position I and III [see inset in panel (e) presenting the position of the {\it d}-block atoms in distorted--kagome lattice].
At bottom panels, the widths of lines are magnified $10$ times.
Meaning of colors and width of lines is the same as in Fig.~\ref{fig.ban191}.
\label{fig.ban189}
}
\end{figure*}

The chiral modes emerge in each studied system, but their frequencies depend strongly on the atomic composition of the compound. 
To describe the interplay between the modes with non-zero chirality and other vibrations, the phonon density of states (DOS) and the atom-projected partial DOS spectra shown in Fig.~\ref{fig.dos191} should be discussed together with the phonon dispersion relations (Fig.~\ref{fig.ban191}).  
The contributions of the vibrational frequencies of {\it p}-block and {\it d}-block atoms to the total DOS are mainly related to the mass ratio between atoms (cf.~Table~\ref{tab.latt}). In the case of systems with a large mass difference, these partial DOSs are well separated, as it is seen for CoSn, FeSn, NiIn, and RhPb, where the phonon modes of {\it p}-block ({\it d}-block) atoms are located at the lower (higher) frequencies. 
In the case of nearly equal-masses (e.g. PtTl), phonons cover the same frequency regions.
The chiral modes contribute to the  partial DOS of the {\it p}-block atoms in the honeycomb sublattice (marked in blue). 
These modes can be derived from local minima and maxima of phonon dispersion relations in high symmetry points (mostly in K, K', and M or related H, H', and L; see Fig.~\ref{fig.ban191}), and are related to the chiral modes propagating along the $c$ direction. 
In the low frequency range, comprising the lowest chiral mode branch, small peaks in DOS corresponding to the chiral modes are observed. 
However, they are not the dominating components of DOS.   
The main contribution of chiral modes is observed in the mid-frequency range, where several peaks related to the Van Hove singularities are located. 
In this part, the vibrations of atoms in a honeycomb sublattice dominate, however they are often mixed with the vibrations of other atoms.
The flat-bands (between K-K' or H-H' points) are also reflected in the DOS in the form of clearly visible peaks.
A good example is the low-energy flat-band corresponding to the vibrations of {\it p}-block atoms in the kagome plane.
For instance, the distinctive peak around $1.75$~THz in Fig.~\ref{fig.dos191}(e).

Contrary to the $ab$ plane, the phonon branches preserve chirality along the $c$ direction (which corresponds to the unchanged color of the line between K and H points in Fig.~\ref{fig.ban191}).
Every wave vector along this path corresponds to the propagation of a chiral mode along the $c$ direction---due to the conservation of angular momentum rule~\cite{tatsumi.kaneko.18}, the mode should be ``topologically'' protected.
If the sample thickness will be smaller than the phonon scattering length but larger than the electron scattering length, the chirality information should not be lost during the propagation~\cite{chen.wu.21}.
In this sense, the chiral modes represent a key step toward utilizing chiral phonons in quantum devices~\cite{li.ren.12}.

The question about the origin of the chiral phonons in this class of systems remains still open~\cite{coh.19}.
Let us start from discussion of the general case.
The lattice dynamics of a system depends on the IFC matrix (including first- and high-order phonon contributions).
In the case of a system with the inversion symmetry, the first order IFC are highly symmetric.
Similarly, when the inversion symmetry is broken, then this breaking will affect the first order IFC matrix.
In this situation, some components of IFC are non-equal under mirror symmetry due to different environments in some sites of the lattice.
As a result, the phonon modes will have angular momentum at a generic non-symmetric point in the Brillouin zone (e.g. this situation is realized in dichalcogenides~\cite{zhu.yi.18,zhang.srivastava.20}).

Moreover, the higher order IFC can play a role in realization of the chiral phonons, especially in system with the broken time reversal (e.g. iron, cobalt, or nickel~\cite{solano.16}).
In the case of CoSn-like compounds, the atoms are located at the high symmetry points, what implicates the preserved inversion symmetry.
In fact, the source of the chiral phonons is ``coded'' in the first order IFC that preserve a three-fold axis of the honeycomb sublattice along the $z$ direction, while the total phonon angular momentum of the system is equal zero~\cite{coh.19}. 
Additionally, in magnetic compounds like FeSn and FeGe the time reversal symmetry can be broken by magnetic moment of iron.
However, the chiral modes are not realized in the layer with {\it d}-elements, due to the triangular lattice.

%%%%%%%%%%%%%%%%%%%%%%%%%%%%%%%%%%%%%%%%%%
%%%%%%%%%%%%%%%%%%%%%%%%%%%%%%%%%%%%%%%%%%

\subsection{CoGe and RhPb in $\bm{P\bar{6}2m}$ symmetry}
\label{sec.sg189}

Our analysis of the systems with the $P6/mmm$ symmetry revealed the soft modes in the phonon dispersion relations of CoGe and RhPb [cf.~Fig.~\ref{fig.ban191}(b) and Fig.~\ref{fig.ban191}(g), respectively].
Thus, both structures are unstable and can be transformed to a distorted structure due to the condensation of the soft phonon mode.
It means that using the displacement pattern attributed to the polarization vectors of this mode, the stable lower-symmetry configurations can be found.
For CoGe and RhPb, the lowest frequency soft modes at the $\Gamma$ point with $B_\text{1u}$ symmetry consist of specific displacements of {\it d}-elements only.
In practice, the triangles constructing the kagome net in $P6/mmm$ crystals rotate approximately by $6\degree$ in opposite directions (see movie \texttt{gamma\_mode4} in SM~\cite{Note1}). 
This soft-mode-induced modification stabilizes the system in the $P\bar{6}2m$ symmetry.

\begin{figure}[!t]
\centering
\includegraphics[width=\linewidth]{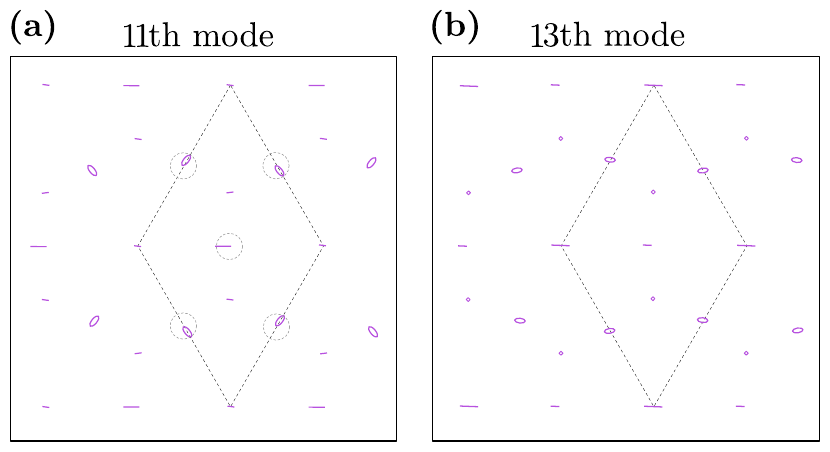}
\caption{
Examples of trajectories (violet lines) of the CoGe atoms (for $P\bar{6}2m$ symmetry), generated by chosen modes at the K/3 point (as labeled).
The dashed rhombus represents the unit cell, while positions of atoms are the same as in Fig.~\ref{fig.traj}.
Here, the gray dashed circles in panel (a) correspond to positions of sites in the ideal kagome lattice.
\label{fig.traj189}
}
\end{figure}

Independently of the positions of {\it d}-elements in the $P\bar{6}2m$ structure, the hexagonal net of {\it p}-element is unchanged and still exhibits the chiral phonons (top panels in Fig.~\ref{fig.ban189}).
The main properties of the chiral phonon branches are similar to those observed in systems with the $P6/mmm$ symmetry [cf.~Fig.~\ref{fig.ban191}(b) and Fig.~\ref{fig.ban191}(g)].
However, as a consequence of the kagome lattice distortion [see the inset in Fig.~\ref{fig.ban189}(e)], the inversion symmetry of the system is lost.
As we discussed in Sec.~\ref{sec.topo_ph}, in such a case the chiral phonons can be expected also in this kagome-like sublattice.
Indeed, the analysis of the phonon angular momentum for {\it d}-element atoms clearly reveals the existence of the chiral modes also in the distorted-kagome lattice.
What is interesting, the chiral modes in this layer are realized only by {\it d}-element atoms, while one {\it p}-element atom shows the ordinary linear movement.

In addition, the {\it d}-element atoms move in an ellipse
(see Fig.~\ref{fig.traj189}) what gives the phonon angular momentum smaller than the nominal.
What is also interesting, the atoms at sites I and II have anti-symmetric position with respect to the {\it p}-element atom in the corner of the unit cell [rhombus in inset of Fig.~\ref{fig.ban189}(e)].
In consequence, similarly like in the honeycomb sublattice, the atoms in these positions have opposite phonon angular momenta.
Also, the atom at position III exhibits in some branches the non-zero angular momentum [bottom panels in Fig.~\ref{fig.ban189}].
However, the total angular momentum of the distorted-kagome lattice is equal to zero.

%%%%%%%%%%%%%%%%%%%%%%%%%%%%%%%%%%%%%%%%%%
%%%%%%%%%%%%%%%%%%%%%%%%%%%%%%%%%%%%%%%%%%

\begin{figure}[!t]
\centering
\includegraphics[width=\linewidth]{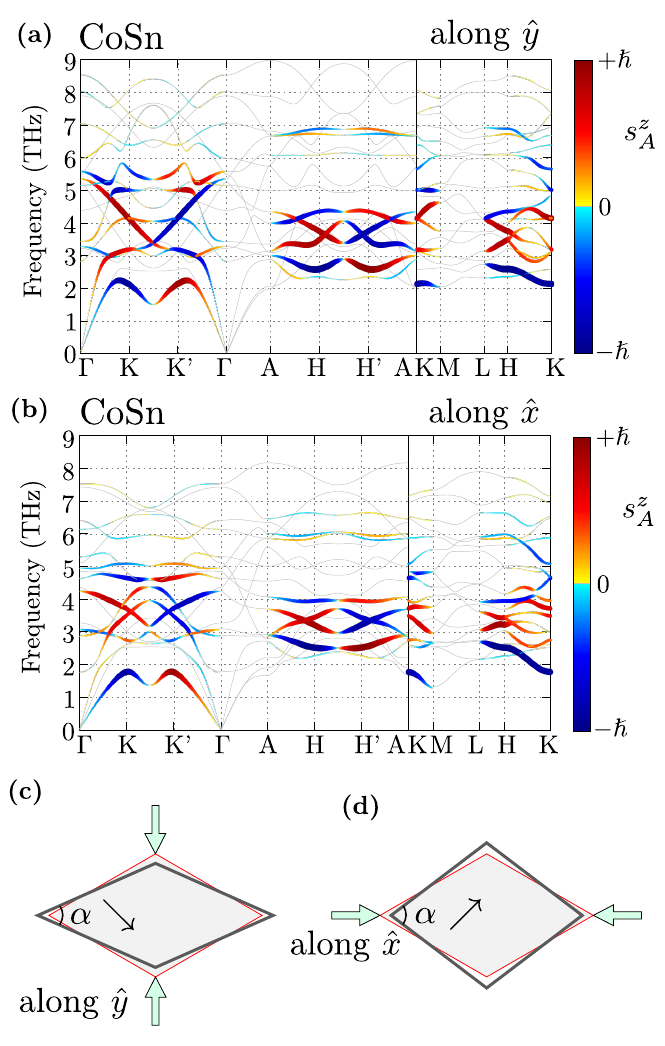}
\caption{
Phonon dispersions of CoSn under uniaxial strain along the $\hat{y}$ and $\hat{x}$ direction, presented in panels (a) and (c), \mbox{or (b) and (d),} respectively.
The meaning of colors and widths of lines are the same as in Fig.~\ref{fig.ban191}.
\label{fig.stres}
}
\end{figure}

\subsection{Influence of uniaxial stress}
\label{sec.stres}

We also examined the effect of the uniaxial stress applied along $\hat{x}$ and $\hat{y}$ directions (Fig.~\ref{fig.stres}). 
Independently of the stress direction, 
the system changes  its symmetry from $P6/mmm$ to $Cmmm$ [mostly due to the modification of the angle between lattice vectors $a$ and $b$ -- cf.~Fig.~\ref{fig.stres}(c) and \ref{fig.stres}(d)].
The first observed consequence is the modification of the 
degeneracy at the $\Gamma$ point. 
Secondly, the crossing of the phonon branches is shifted from the BZ surface or high symmetry point to the interior of the BZ.
Last, but not least, the initially flat-bands change their flexure.
Notably, the main features associated with the chiral phonons are unchanged [cf.\ Figs.~\ref{fig.ban191}(a) and Fig.~\ref{fig.stres}].

Effects induced by the uniaxial stress are similar to those resulting from the magnetic field and spin--orbit coupling modifying the electronic band structure.
In the former case, points which were initially degenerate at the $\Gamma$ point become decoupled.
A similar effect was observed in the case of the acoustic waves~\cite{brendel.peano.17}.
Lifting of the degeneracy leads to the emergence of chiral phonons at the $\Gamma$ point and can be useful for potential experimental detection, e.g.\ by circularly-polarized Raman scattering~\cite{zhang.mao.17} (cf.\ Sec.~\ref{sec.exp}).
Additionally, mixing of the branches around K and K' is observed, what should correspond to the changes induced by the finite spin--orbit coupling, in analogy to the electronic systems.

%%%%%%%%%%%%%%%%%%%%%%%%%%%%%%%%%%%%%%%%%%
%%%%%%%%%%%%%%%%%%%%%%%%%%%%%%%%%%%%%%%%%%
%%%%%%%%%%%%%%%%%%%%%%%%%%%%%%%%%%%%%%%%%%
%%%%%%%%%%%%%%%%%%%%%%%%%%%%%%%%%%%%%%%%%%

\section{Proposed experimental detection}
\label{sec.exp}

As we mentioned above, the chiral phonons in the case of $P6/mmm$ crystals are realized mostly within the honeycomb sublattice, and are composed of {\it p}-element atoms on $2d$ Wyckoff position. 
For these positions, we can find two infrared (IR) active modes $A_\text{2u}$ and $E_\text{1u}$ and one Raman active mode $E_\text{2g}$.
Consequently, chiral modes could be measured in relatively simple way by IR or Raman spectroscopy.
The realization of chiral phonons at the $\Gamma$ point under uniaxial pressure allows for performing these types of experiments.

Additionally, a phonon with nonzero angular momentum can be created by emission or absorption of a photon with a given circular polarization~\cite{zhang.niu.15,chen.wu.21}.
During this process, the intervalley electron scattering at valley centers involving a valley center phonon is expected.
However, to improve the description of these processes, a study of the valley structure in electronic bands is needed, which is out of the scope of this paper.
In this type of studies, the circular-polarized Raman spectroscopy can be employed -- this allows for the helically resolved spectroscopy~\cite{chen.zheng.15}, and to study modes with different types of chirality.

Recent experiments suggest the phononic-to-electronic conversion of angular momentum~\cite{ruckriegel.streib.20,hamada.murakami.20b}.
For example, the phonon angular momentum of the surface acoustic wave can control the magnetization of a ferromagnetic layer, what was shown in the case of Ni film in Ni/LiNbO$_{3}$ hybrid device~\cite{sasaki.nii.20}.
Similarly, the external magnetic field can lead to a decoupling of the frequencies of the initially degenerate modes~\cite{schaack.76}.
Indeed, recent experiments confirm this expectation~\cite{juraschek.narang.20}.
In such a way, modification of the phonon spectra at the $\Gamma$ point under external magnetic field should be observed.
Equivalently, the uniaxial pressure can be applied resulting in a similar effect (what was discussed in Sec.~\ref{sec.stres}).

The emergence of a nonzero phonon angular momentum within the hexagonal sublattice of CoSn-like compounds, can be interesting also from the experimental point of view. 
Similarly to the electronic systems~\cite{zhang.tan.05}, in a properly prepared sample (in slab form), phononic angular momentum Hall effect could be realized~\cite{park.yang.20} (in analogy to the integer quantum Hall effect).
Similar effect was predicted for the photonic chiral edge state propagating along a surface~\cite{peng.yan.20}.
Realization of the edge mode with well defined chirality can be induced by photons with the same circular polarization~\cite{wu.hu.15}.
Additionally, the occurrence of such edge modes should lead to a modification of the transport properties of such systems~\cite{xiao.yao.07}.

\section{Summary}
\label{sec.sum}

Our studies confirm that
the ideal honeycomb~\cite{zhang.niu.15} and kagome~\cite{chen.wu.19} lattices allow for the realization of chiral phonons (with nonzero angular momentum). 
Typically, the chiral phonons are observed in two dimensional materials~\cite{chen.zhang.18} and might be an interesting addition to the quantum devices based on phonons~\cite{li.ren.12}.
In this paper we discussed the realization of chiral phonons in the CoSn-like compounds--- layered systems build by decorated kagome and honeycomb lattices.
We demonstrated that the realization of an ideal honeycomb lattice by the {\it p}-block elements allows for the emergence of chiral phonons within this sublattice.
In our study of CoSn-like compounds, based on the first principles method, we showed that the properties of the chiral phonons do not depend on the mass ratio of constituent elements or on their intrinsic magnetic order.

Typically, CoSn-like compounds crystallize in the $P6/mmm$ structure. However, 
in two of the discussed compounds, i.e., in CoGe and RhPb, soft modes were revealed. 
We found that these soft modes lead to the stable $P\bar{6}2m$ structure.
In this case, the absence of the inversion symmetry allows for the realization of 
the chiral modes also by the {\it d}-block atoms within distorted-kagome sublattice.
We remark that the symmetry of the system can be also changed by an uniaxial strain.
The strain effect can mimic the magnetic field or spin--orbit coupling 
acting on phonons.
Regardless of the modification of the symmetry, the main 
character of the chiral phonons remains unchanged.
Additionally, we proposed an experimental method for confirmation of the described here chiral phonons in CoSn-like compounds.

\begin{acknowledgments}
Some figures and movies in this work were rendered using {\sc Vesta}~\cite{momma.izumi.11} and {\sc VMD}~\cite{vmd} software.
This work was supported by 
National Science Centre (NCN, Poland) under Projects No.
2016/21/D/ST3/03385 (A.P.), 
2018/31/N/ST3/01746 (A.K.), 
2016/23/B/ST3/00839 (A.M.O.),
and
2017/25/B/ST3/02586 (P.P.). 
A.P. \mbox{appreciates} funding in the frame of scholarships of the 
Minister of Science and Higher Education (Poland) for outstanding young 
scientists (2019 edition, No. 818/STYP/14/2019).
\end{acknowledgments}

%%%%%%%%%%%%%%%%%%%%%%%%%%%%%%%%%%%%%%%%%%
%%%%%%%%%%%%%%%%%%%%%%%%%%%%%%%%%%%%%%%%%%
%%%%%%%%%%%%%%%%%%%%%%%%%%%%%%%%%%%%%%%%%%
%%%%%%%%%%%%%%%%%%%%%%%%%%%%%%%%%%%%%%%%%%

\bibliography{biblio.bib}

\end{document}